# Predicting polarisation hemisphere switch in $C_{60}$ caused by the motion of an internal point charge with an electrostatic and quantum chemistry solutions

G. Raggi

*School of Physics, University of Nottingham, University Park, Nottingham, NG7 2RD, UK. E-mail: Gerardo.Raggi@nottingham.ac.uk.*

**A classical electrostatic solution for polarisation charge on the interface of a dielectric cavity interacting with an internal point charge is presented. This solution comes from the Gauss electrostatic potential as a sum of two terms, the cavity and the point charge, expanded with Legendre polynomials. Subsequent application of the Dietrich-Newman boundary conditions defines the problem of emptiness in the inside and isotropic dielectric medium on the outside of the spherical interface to obtain an equation which describes the surface charge density on the interface. These results are compared with quantum chemical calculations using density functional theory for neutral $C_{60}$ fullerenes. Comparison showed that there was good qualitative agreement between the classical electrostatic theory and quantum calculations. The polarisation effect that occurs as a result of the motion of the trapped particle inside the $C_{60}$ molecule shows potential for a polarisable nanoswitch which might be used in nanotechnology as electronic component.**

Since Lord Kelvin in 1850[1,2] a huge amount of work has been done for conductors and for dielectrics since Kirkwood[3] in the 30's especially the electrical contribution for the chemical potential of an ion using Debye-Hünckel theory[4] for a point charge inside of a dielectric sphere. More recently, in 1992 the image theory method was applied for dielectrics, particularly for a homogeneous and layered dielectric sphere by Lindell *et al*.[5,6] Other analytic expressions were obtained by Ohshima in the form of an explicit infinite series for two dissimilar hard and soft (ion-penetrable) spheres with constant surface charge density.[7–9] In the same dielectric perspective Cai *et al* proposed a method to extend the fast multipole method (FMM) based on a Green's function for the Laplace operator or the Helmholtz operator to calculate the electrostatic potential due to charges inside and outside of a dielectric sphere.[10] Bichoutskaia *et al* proposed a model between two dissimilar dielectric spherical particles obtained from Gauss's law that couples the distribution of polarisation charge on the surfaces.[11]

In previous publications, an analytical solution for the distribution of charge on a dielectric sphere due to the presence of an external point charge has been presented. This describes how charge on the surface is polarised by an electric field into regions of negative and positive charge.[12] That solution is based on the assumption that the polarisable charge resides on the interface of the dielectric sphere and no polarisation charge is present inside of the dielectric sphere, and the outside of the sphere is vacuum. Most of the electrostatic solutions for a point charge and a particle on both conductors and dielectrics are for the point charge being located outside of the sphere. One of few solutions for a particle inside a dielectric was proposed by Sten *et al*[13] in complement of a previous theory[5] handled by infinite series of spherical harmonics and solved in terms of image line and point sources on a point charge located outside and inside. The sphere in both cases is filled with a dielectric material, this prior work cannot be compared with the present paper because the model involves a hollow sphere with a dielectric material surrounding it. Cases in which the charge is inside of the sphere have significant relevance in biological structural analysis specifically to physical applications.[14] Other physical applications of charge residing in the inside cage molecule has been used for molecular switching purposes such as Ca@$C_{60}$ and Li@$C_{60}$ endofullerenes based on charge-transfer excitation.[15,16]

When a particle is moving outside of the $C_{60}$ cage, the polarisation affects the regions of the $C_{60}$ according to a classic electrostatic model[11,12,17]. In this particular case an analytical solution of polarisation charge in the spherical interface due to the presence of an internal point charge: a classical solution for a particle moving inside of a spherical cavity cage with the same characteristics of a $C_{60}$ molecule was used in order to compare with quantum calculations on $C_{60}$ fullerene assuming that the partial atomic charges reside at the interface between the internal wall of the molecule and its internal empty space. The polarisation is a direct consequence of the distribution of charge induced by the presence of an electric field; in this case, as well, the polarisation is induced on the spherical interface by the presence of the point charge inside of the sphere. This surface charge density $\sigma(s)$ is related by the Gauss electrostatic potential $\Phi(r)$ as follows:

$$\Phi(\boldsymbol{r}) = K \int \frac{\sigma(s)ds}{|\boldsymbol{r}-\boldsymbol{a}|} + K\frac{q}{|\boldsymbol{r}-\boldsymbol{h}|} \qquad (1)$$

Equation (1) is the sum of the contributions of the spherical interface surface charge (first term) and the point charge (second term) to the potential $\Phi(r)$ where $a$ represents the radius of the sphere, $h$ the distance from the point charge to the centre of the sphere (see Figure 1), $q$ the magnitude charge of the point charge and $K = 1/4\pi\varepsilon_0$ ($\varepsilon_0 = 8.85 \times 10^{-12}\ F\ m^{-1}$) as a proportionality constant. For the spherical surface charge represented by the first term of equation (1), there is a discontinuity when $r \equiv a$ which leads to two separate equations of the potential $\Phi(r)_{r<a}$ when $r < a$ and $\Phi(r)_{r>a}$ when $r > a$. Using a Legendre polynomial expansion for the term $1/|r - a|$ of equation (1) states

$$\frac{1}{|r-a|} = \sum_{l=0}^{\infty} \frac{r^l}{a^{l+1}} P_l(\cos\beta) P_l(\cos\theta) \qquad (2)$$

Substitution of above equation into the first term of equation (1) and subsequent integration over the surface in spherical polar coordinates $(r, \theta, \varphi)$ for vector $r$ and $(a, \beta, \varphi)$ for vector $a$ from $0 < \varphi < 2\pi$ takes the form

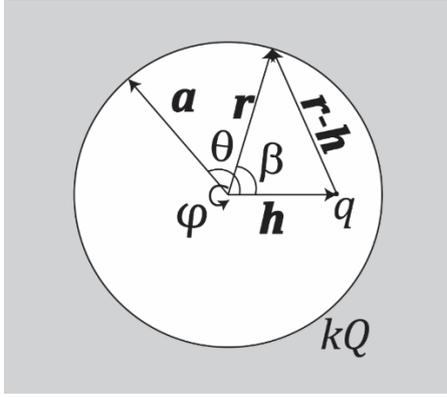

Figure 1 Geometric representation of the spherical interface of the cavity and internal point charge.

$$K \int \frac{\sigma(\theta) a^2 \sin\theta\, d\theta d\varphi}{|r-a|} = \sum_{l=0}^{\infty} A_l \frac{r^l}{a^{2l+1}} P_l(\cos\beta) \qquad (3)$$

where

$$A_l = 2\pi K \int_0^\pi \sigma(\theta) a^{l+2} \sin\theta\, P_l(\cos\theta) d\theta \qquad (4)$$

The second term of equation (1) represents the internal point charge and there is no discontinuity, which means it is the same equation for both solutions $r > a$ and $r < a$

$$K \frac{q}{|r-h|} = qK \sum_{l=0}^{\infty} \frac{h^l}{r^{l+1}} P_l(\cos\beta) \qquad (5)$$

Substitution of eq. (3) and (5) into (1) gives the expansion of the Gauss potential in terms of Legendre polynomials for $r < a$

$$\Phi_{r<a} = \sum_{l=0}^{\infty} A_l \frac{r^l}{a^{2l+1}} P_l(\cos\beta) + qK \sum_{l=0}^{\infty} \frac{h^l}{r^{l+1}} P_l(\cos\beta) \qquad (6)$$

and for $r > a$ again substituting (3) and (5) into (1)

$$\Phi_{r>a} = \sum_{l=0}^{\infty} A_l \frac{1}{r^{l+1}} P_l(\cos\beta) + qK \sum_{l=0}^{\infty} \frac{h^l}{r^{l+1}} P_l(\cos\beta) \qquad (7)$$

Imposing the Dietrich-Newman boundary conditions and assuming the dielectric medium is isotropic, and the normal components of the dielectric displacement field **D** and the tangential components of the electric field **E** satisfy the following equations[11,18]

$$(\mathbf{D}|_{r=a^+} - \mathbf{D}|_{r=a^-}) \cdot \mathbf{n} = \sigma \qquad (8)$$

and

$$(\mathbf{E}|_{r=a^+} - \mathbf{E}|_{r=a^-}) \times \mathbf{n} = 0 \qquad (9)$$

where $a^+$ defines the radial position approaching from outside, and $a^-$ on the inside of the sphere; **n** is a unit vector normal to the surface and directed outwards from the spherical interface and $\sigma$ the surface charge density. Both equations come directly from Maxwell equations.[18] At the interface the normal component of the electric field is discontinuous due to the presence of a permanent surface charge, noting that $\mathbf{D} = \varepsilon_0 \mathbf{E}$ (for vacuum) and $\mathbf{E} = -\nabla\Phi$ therefore from equation (8)[12]

$$\frac{\sigma_t}{\varepsilon_0} = \left.\frac{\partial \Phi_{r<a}}{\partial r}\right|_{r=a^-} - \left.\frac{\partial \Phi_{r>a}}{\partial r}\right|_{r=a^+} \qquad (10)$$

Differentiating equation (6) and equation (7) and substituting in (10) the equation for charge density becomes

$$\sigma_t(\cos\beta) = \frac{1}{4\pi K} \sum_{l=0}^{\infty} A_l \frac{2l+1}{a^{l+2}} P_l(\cos\beta) \qquad (11)$$

The other boundary condition is related to the excess or free charge where the electric field is reduced in the exterior of the sphere $r > a$, by a factor of the dielectric constant $k$. This reduction can be understood in terms of polarisation, such as the reduction of the electric field in opposition to that of the internal charge. This boundary condition states that the normal component of the dielectric displacement field due to the presence of an excess of charge on the surface of the sphere satisfies the following equation

$$\frac{\sigma_e}{\varepsilon_0} = \left.\frac{\partial \Phi_{r<a}}{\partial r}\right|_{r=a^-} - k\left.\frac{\partial \Phi_{r>a}}{\partial r}\right|_{r=a^+} \qquad (12)$$

Differentiating equations (6) and (7) and substituting in (12)

$$\frac{\sigma_e}{\varepsilon_0} = \sum_{l=0}^{\infty} A_l \frac{k(l+1)+l}{a^{l+2}} P_l(\cos\beta) + (k-1)qK \sum_{l=0}^{\infty} \frac{(l+1)h^l}{a^{l+2}} P_l(\cos\beta) \qquad (13)$$

where $k = \varepsilon/\varepsilon_0$ is the dielectric constant. Multiplying both sides of equation (13) by $\sin\beta\, P_{l'}(\cos\beta)$, integrating over the surface of the sphere and using the following relation

$$\int_0^\pi \sin\beta\, P_{l'}(\cos\beta) P_l(\cos\beta) d\beta = \frac{2\delta_{l,l'}}{(2l+1)} \quad (14)$$

with $l' = 0$ the left side of equation (14) becomes

$$\int_0^\pi \sin\beta\, P_0(\cos\beta) P_l(\cos\beta) d\beta \frac{\sigma_e}{\varepsilon_0} = 8\pi K \sigma_e \delta_{l,0} \quad (15)$$

while the right side of equation (13) becomes

$$\cdots = \frac{2}{a^2}\left(A_l \frac{k(l+1)+l}{(2l+1)a^l} + (k-1)qK \frac{l+1}{2l+1}\frac{h^l}{a^{l-1}}\right) \quad (16)$$

Multiplying right sides of (15) and (16) by $\frac{a^2}{2}$, equation (13) takes the form

$$4\pi K a^2 \sigma_e \delta_{l,0} = A_l \frac{k(l+1)+l}{(2l+1)a^l} + (k-1)qK\frac{l+1}{2l+1}\frac{h^l}{a^{l-1}} = Kq_e \delta_{l,0} \quad (17)$$

where the excess of charge density is the total charge divided by the surface area $\sigma_e = \frac{q_e}{4\pi a^2}$ in the sphere[11]. From (17)

$$A_l = \frac{(2l+1)a^l K q_e}{k(l+1)+l}\delta_{l,0} + qK\frac{(1-k)(l+1)}{k(l+1)+l}ah^l \quad (18)$$

and substituting in (11)

$$\sigma_{pol}^{surf}(\cos\beta) = \frac{q_e}{4\pi k a^2} + \frac{1}{4\pi a}\sum_{l=0}^\infty q\frac{(1-k)(l+1)(2l+1)}{k(l+1)+l}\left(\frac{h}{a}\right)^l P_l(\cos\beta) \quad (19)$$

This equation describes the surface charge density on the spherical interface, with radius $a$ and dielectric constant $k$, due to the presence of a point charge inside of an empty cavity located at a distance $h$ from the centre of the cavity with charge $q$.

Figure 2 illustrates the surface charge density when the point charge is moving inside and on the left hemisphere of a neutral sphere (see Figure 3a to Figure 3d) using equation (19) with radius $a = 3.8$ Å and dielectric constant $k = 3.45$ which has the dimensions and the dielectric properties of a $C_{60}$ molecule.[19,20] Since it is a neutral sphere the excess of charge $q_e = 0$ and the charge of the point charge $q = +2e$. The values that the distance $h$ takes, for the point charge when it is moving on the left hemisphere, are from -3.0 Å to 0.0 Å in steps of 0.2 Å. As the sphere has symmetrical symmetry, the differential surface charge density when the point charge is moving on the right hemisphere would have mirror curves with respect to $\beta$ angle, that means the lobes of negative charge of the graphic would be on the left side about zero, and the positive charge would be on the right side. At the distance of -3.0 Å from the centre (only 0.8 Å from the shell) the negative charge is concentrated between 2.4 and π equivalent to ~13%[ξ] of the surface area and the positive is on the remaining ~87% of the surface. While the particle is moving away from the wall or shell to the centre, the negative charge spreads over the surface until -0.0001 Å; at this point half of the hemisphere (or 50% of surface charge) is purely negative and the other half purely positive (see inset of Figure 2). The same happens symmetrically in the other side of the hemisphere from 0.0001 Å to 3.0 Å.

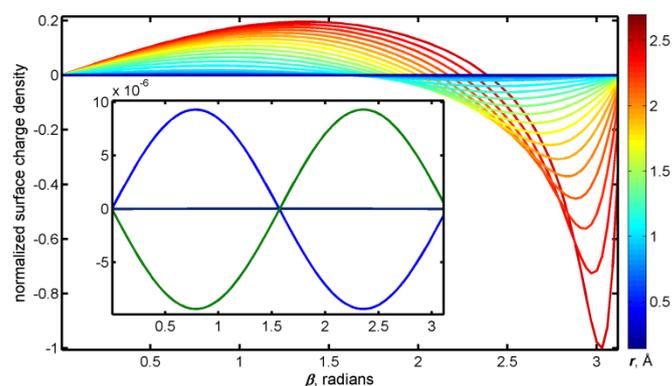

Figure 2 Normalized surface charge density, calculated as $2\pi a^2 \sin(\beta)\, \sigma(\cos\beta)$ using equation (19) on a sphere representing the dimensions and dielectric properties of the $C_{60}$ molecule: $a = 3.8$ Å, $k = 3.45$ and $q_e = 0$. The point charge has a magnitude of $q = +2e$. The colour labelling is used to represent the separation $h$, between the centre of the sphere and the point charge which is moving in the left hemisphere only, $h$ varies with a step of 0.2 Å in the distance range of -3.0 Å (dark red) to 0 Å (dark blue). Inset shows the surface charge density distribution at $h = -0.0001$ Å (light blue) when the point charge is in the left hemisphere, $h = 0$ Å (flat dark blue line) when the point charge is at the centre) and $h = 0.0001$ Å (green) when the point charge in the right hemisphere.

A switch of polarisation, from negative charge to positive charge, between hemispheres was observed when the point charge moved from one side of the fullerene to the other in a short distance range away from the middle of the sphere (-0.0001 Å to 0.0001 Å). This switch is depicted in the inset of Figure 2 for three positions of the point charge when the point charge is at -0.0001 Å (dark blue) the lobe of negative charge represents the charge density of the left hemisphere and the lobe of positive charge represent the charge density of the right hemisphere (see Figure 3c). When the particle is in the middle of the sphere $h = 0$ the charge density is represented by the flat line in black at 0, which means the shell is now neutral and no polarisation is observed. Just crossing the zero on the right hemisphere at 2.0 Å, given the symmetry, the left hemisphere becomes positive and the right hemisphere becomes negative represented by the green line in the inset.

In the next section these electrostatic results of the surface charge density for empty sphere and an internal point charge are compared with quantum chemical calculations of $C_{60}$ simulating the sphere of the electrostatic model in interaction with an internal point charge to be incorporated into the calculations.

Quantum calculations using density functional theory (DFT) have been carried out to compare with the classical electrostatic calculations of the previous section. The distribution of charge in the $C_{60}$ was obtained with the Becke three parameter hybrid exchange functional and Yang's gradient B3LYP[21,22]/6-311G*[23] level of theory, implemented in the *Q-Chem* quantum chemistry package.[24] The charge density is taken from partial atomic charges from Mulliken population analysis[25] and compared with the classical results from equation (19). The partial atomic charges of Mulliken population analysis show a high degree of sensitivity to the basis set.[26,27] However, using a consisting basis set proved a high degree of quality confidence in previous publications.[12,15,28] These values of partial atomic charges of Mulliken population analysis are depicted on Figure 3 for six values of separation distance between the centre of the $C_{60}$ and the point charge. At a distance of -3.0 Å (Figure 3a) the negative charge is concentrated on ~12% of the surface area of the fullerene and the positive charge is smeared over the rest ~88% of the surface area. The small difference of 1% between this value of surface area and the one calculated with the electrostatic model (13%) for the same distance is perhaps caused by the nature of the electrostatic model that is a continuous model and the quantum model which is a discrete model; one is provided by an analytical solution and the other by overlapping electron orbitals into the net population analysis respectively. At -0.8 Å (Figure 3b) of separation from the centre of the surface, the area covered by the negative charge is about ~33%, instead the classical solution has a value of 44%, which is 11% different to the quantum solution. When the internal charge is moving towards the centre, at -0.2 Å and 0.2 Å the negative charge spreads over the surface on about ~50% until it reaches the centre, where the negative and positive charges are equally spread on a non-polarized fullerene.

The method of selection of partial atomic charges from Mulliken population to compare with classical electrostatic comes from dividing by rings over the fullerene and make a summation of Mulliken charges in the ring. The rings are defined between two angles $\beta_i$ and $\beta_{i+1}$ by tracing a line from the centre of the fullerene and one point over the surface on incremental steps $\Delta\beta = \beta_{i+1} - \beta_i$. This angle must be chosen with two restrictions: avoid missing any surface area or atoms *i.e.* multiple of $\pi$ radians and it also must be big enough to contain at least one atom. The angle chosen to compare the fullerene and the particle inside was $\Delta\beta = 20°$, this gives nine slices of the surface of $C_{60}$ with the same width. Within these slices summations of the partial atomic charge has been made, and this has been depicted in Figure 4 as a blue colour line fitted with a polynomial of degree 8.

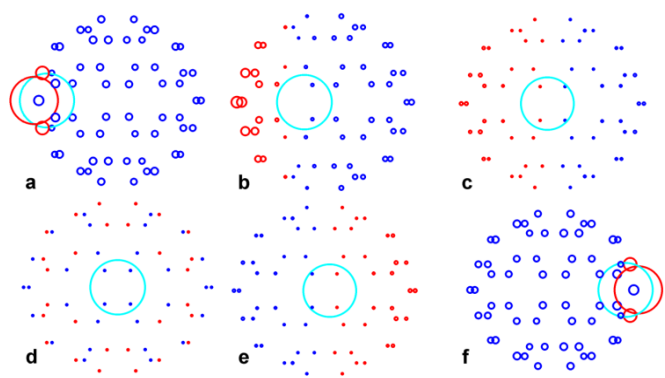

Figure 3 Distribution of partial atomic charges on the $C_{60}$ molecule at six values of separation distance between the centre of $C_{60}$ and the internal point charge of $q = +2e$ located: (a) -3.0 Å, (b) -0.8 Å, (c) -0.2 Å, (d) 0 Å, (e) 0.2 Å and (f) 3.0 Å. Blue circles show positive charge residing on carbon atoms, red circles depict negative charge and green circle the internal point charge. The size of each circle has been renormalized to convey the degree of polarisation rather than its magnitude.

The electrostatic result taken by the analytical solution equation (19) is represented by the cyan colour line in Figure 4, calculated as $2\pi a^2 \sin(\beta)\,\sigma(\cos\beta)$ on a sphere with the dimension ($a$ = 3.8 Å) and dielectric properties ($k$ = 3.45) of the $C_{60}$ molecule, these classical results were scaled in order to compare with the atomic charge density results of the Mulliken population analysis. At a distance of -3.0 Å from the centre (Figure 4a) the negative charge is concentrated in terms of the angle $\beta$ between 2.54 and 3.14 rad with a magnitude of ~1.5 a.u. due to the proximity of the point charge to the $C_{60}$ which is in a good agreement with the electrostatic model. When the particle is closer to the centre at -0.8 Å or 0.8 Å (Figure 4b) the quantum solution crosses the x-axis at 1.86 rad while the classical at 1.75 rad giving a little difference between them but whit almost the same area covered. At -0.2 Å (Figure 4c) the positive and negative charge density are separated by hemispheres (see Figure 3c and Figure 3e) and the lines crosses the x-axis at the same point (1.6 rad); however, the charge magnitude is 3.5 times smaller (0.04 a.u.) than that when it was at 0.8 Å (0.14 a.u.). Since it is a neutral sphere the charge density is zero on all points over the sphere when the particle is in the centre and the analytical solution which is a continuous model is represented by the flat line at zero in Figure 4d which is contrasted by the curves showing equal distributions of positive and negative charge with small magnitudes of partial atomic charge around the order of $10^{-3}$ a.u.

Polarisation can affect the interaction between dielectric particles[29–34] or can define a key element in electronic components.[35–37] Polarisation for this particular case, where the point charge is located on the inside of the particle might be used as a mechanism for switching charge between stable states with the simple movement of a component (*i.e.* an atom) of the particle. If this component is trapped inside of the particle the degrees of freedom are reduced and it is easily controlled (mainly because a finite number of stable states exists). The way to control it could be via external stimuli such as light, temperature, voltage, electric or magnetic fields. For $C_{60}$ the position of the charged particle on the inside of the neutral sphere drastically changes the polarisation on the surface of the sphere and this change is the key to defining a nanoswitch. The Ion can be substitute by an atom in which depending on the position might transfer charge to the shell or not.[15,38]

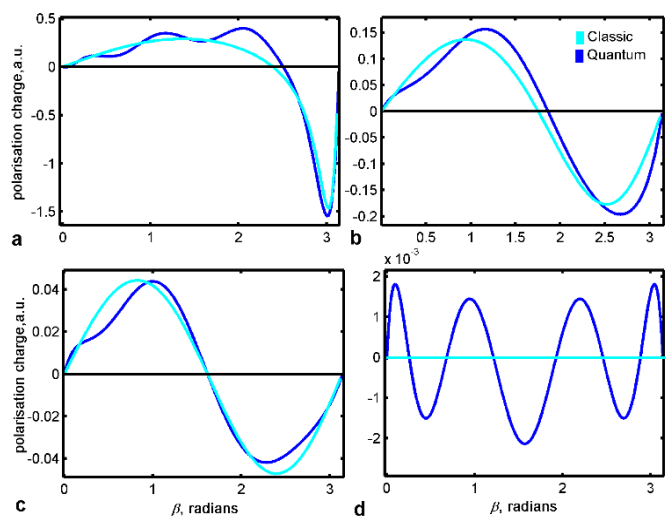

Figure 4 Surface charge density of a sphere in the presence of an external point charge. The cyan line represents the analytical solution calculated as $2\pi a^2 \sin(\beta)\,\sigma(\cos\beta)$ using equation (19) on a sphere representing the dimensions and dielectric properties of the $C_{60}$ molecule: a = 3.8 Å, $k$ = 3.45 and $q_e = 0$ normalized to the quantum charge density results. The quantum charge density results are depicted by the blue line of $C_{60}$ molecule taken from Mulliken population analysis when the inside point charge is at the positions (a) -3.0 Å, (b) -0.8 Å, (c) -0.2 Å and (d) 0.0 Å.

According to the classical model elaborated for an empty cavity surrounded by dielectric material hollow sphere interacting with an internal point charge a switch of polarisation was created, given by a redistribution of charge by crossing the centre of the fullerene from one of the hemispheres to the other. The results of this analysis have been compared with quantum chemical calculations to obtain a qualitative description of charge distribution. Both analyses have good qualitative similarities with respect to the distribution of charge and the small differences between these two methods could be explained by the scale in which the quantum effect seems to take control in certain positions of the internal particle. This polarisation on the surface of the sphere or fullerene can be used to define a switch through the motion of an ion or charged particle in the cavity by restricting the movement of the trapped ion in a path on the centre of the cavity from one hemisphere to the other. This restriction on the movement can be achieved by electrical, magnetic or optical means applied to the particle.[38]

## Notes and references

$^\xi$ Calculated as $\int_0^{2\pi}\int_{\theta_1}^{\theta_2} r^2 \sin\theta\, d\theta\, d\varphi / 4\pi r^2$

## Acknowledgements

The author wishes to thank Professor T. Wright, Professor A. J. Stace, Dr H.K. Chan and Professor E. Besley for fruitful discussions and comments, Dr T. Fernholz and Dr C. H. Lam for financial support and finally to the High-Performance Computing (HPC) Facility at the University of Nottingham for providing computational time.